\documentclass[aps,prl,twocolumn,superscriptaddress,showpacs]{revtex4}
\usepackage{natbib}
\usepackage[dvips]{graphicx}

\begin{document}

\title{Parity Effects in Spin Decoherence}

\author{A. Melikidze}
\affiliation{Kavli Institute for Theoretical Physics,
  University of California, Santa Barbara CA 93106, USA}

\author{V. V. Dobrovitski}
\affiliation{Ames Laboratory, Iowa State University, Ames IA 50011, USA}

\author{H. A. De Raedt}
\affiliation{
  Applied Physics - Computational Physics, Materials Science Centre,
  University of Groningen, Nijenborgh 4, NL-9747 AG
  Groningen, The Netherlands}

\author{M. I. Katsnelson}
\affiliation{Uppsala University, Department of Physics, SE-751 21,
  Uppsala, Sweden}

\author{B. N. Harmon}
\affiliation{Ames Laboratory, Iowa State University, Ames IA 50011, USA}

\date{April 14, 2004}

\begin{abstract}

We demonstrate that decoherence of many-spin systems can drastically
differ from decoherence of single-spin systems.  The difference
originates at the most basic level, being determined by parity of the
central system, i.e. by whether the system comprises even or odd
number of spin-1/2 entities. Therefore, it is very likely that similar
distinction between the central spin systems of even and odd parity is
important in many other situations. Our consideration clarifies the
physical origin of the unusual two-step decoherence found previously
in the two-spin systems.

\end{abstract}


\maketitle


\section{Introduction}

Reduced dynamics of a small quantum system coupled to a bigger
environment has recently become the subject of particularly active
investigation.  In fields like quantum optics~\cite{Mandel_Wolf} and
quantum computation,~\cite{Nielsen_Chuang}, there is a naturally
defined distinct ``central'' system (i.e. an atom or a qubit) which
interacts with its environment, and whose dynamics is of primary
importance.  Similar situations are often encountered in the condensed
matter physics, e.g., when considering a heavy particle tunneling in a
crystal, tunneling centers in glasses~\cite{TLS}, Kondo systems
~\cite{Hewson} etc.  This problem is also of importance when a
naturally defined central system is absent, such as in a recently
developed promising approach to the theory of strongly correlated
systems, the dynamical mean-field theory (for review see
~\cite{DMFT}). In this approach, the system of interacting particles in
a crystal is replaced by an ``effective impurity'' in a
self-consistently defined thermostat.

So far, quantum evolution of a single two-level system (or,
equivalently, a single spin-1/2 entity) interacting with a bath of
bosons~\cite{TLS} or spins~\cite{Garg,Prokof'ev_Stamp} has been studied in
much detail. In contrast, the central systems comprising several
strongly interacting spins 1/2 have not been that extensively
investigated. A general analysis of the two-spin central system
interacting with a bath of bosons has been presented in
Ref.~\cite{DubeStamp}, but more detailed considerations are
lacking. Several interesting cases of a two-spin system coupled to a
spin bath have been considered in Refs.~\cite{DobRaeKanHar,ourkondo},
and it has been demonstrated that behavior of many-spin central
systems can be very different from a single-spin case.  Consideration
of many-spin central systems is of particular importance for possible
implementation of quantum computations which use several strongly
coupled two-level systems for encoding of a single qubit
~\cite{spin2,spin3}.  This
representation allows using the ``decoherence-free subspaces'' and
error-correcting schemes developed for multi-spin qubits
~\cite{dfs,errorcor}.

In this work, based on an exactly solvable but realistic model, we
show explicitly that decoherence of a two-spin-1/2 system can be
qualitatively different from decoherence of a single spin 1/2. We
demonstrate that this difference originates at the most basic level,
and is determined primarily by parity of a central system, i.e. by
whether the central system comprises even or odd number of spin-1/2
entities.

It is known that the parity of the spin system is the cause of the
drastically different behavior in the tunneling of magnetization in
a wide class of spin systems such as magnetic nanoparticles and
molecular magnets where the tunneling is due to magnetic anisotropy
or magnetic field \cite{Loss-DiVincezo,Delft-Henley}. In this paper, we
explore a different effect, in which the parity of the central system
determines the long-time dynamics of the decoherence process.
We emphasize that in the system considered here the
quantum oscillations are caused by the isotropic exchange interaction
and are independent of the symmetry of the crystal field and external
magnetic field; thus the short-time oscillations
do not depend on the parity.


Although there are many possible central systems coupled to various
kinds of spin baths, the generic differences between the many-spin and
the single-spin central systems can be understood based on simple
models.  An instructive
model of a many-spin central system interacting with a spin bath, has been
recently analyzed by Dobrovitski et al.~\cite{DobRaeKanHar} This model
is aimed to describe (at least, qualitatively) main features of such central
systems as magnetic molecules, quantum dots or impurity spins 
which experience decoherence from
the nuclear spin bath. In these systems, the dominant interaction with
the nuclear spins can be approximated by the isotropic Heisenberg
interaction, since anisotropic interactions are often small.
The model
is defined by the Hamiltonian:
\begin{equation}
\label{Model}
H = \vec C^2 + 2\vec C\sum\limits_{k=1}^N J_k\vec s_k,
\end{equation}
which describes the central system composed of two spins: $\vec C =
\vec c_1 + \vec c_2$, $c_1=c_2=1/2$, which is coupled by Heisenberg
exchange interaction to $N$ environmental spins $s_k=1/2$, $k=1\ldots
N$. Note that the environmental spins don't have their own dynamics.
This may be viewed as a limit case where the dynamics of the central
system is much faster than that of the environment.

A special feature
of this model, which makes it different from the ``central spin''
models considered by Garg, or Prokof'ev and
Stamp,~\cite{Garg, Prokof'ev_Stamp} is
the fact that in our treatment the central system is not reduced to the
doublet of lowest states. This features is crucial to the results
discussed below.

One is interested
in the time-evolution of the initial system-plus-environment state
which is taken in the form:
\begin{equation}
\label{Initial_condition}
|i\rangle = |\uparrow\rangle_{c_1}|\downarrow\rangle_{c_2}
\prod\limits_{k=1}^N |i\rangle_{s_k}.
\end{equation}
The initial states of the environmental spins $|i\rangle_{s_k}$ are
assumed random and uncorrelated. The initial state of the system
is a superposition of the singlet and triplet states of the two central
spins:
\begin{equation}
|\uparrow\rangle_{c_1}|\downarrow\rangle_{c_2}
  =  \frac {1}{\sqrt 2} \left( |1,0\rangle_C + |0,0\rangle_C \right),
\end{equation}
where we have introduced notation $|C,C^z\rangle_C$ for the central spin.
One considers the problem of the decay of this coherent singlet-triplet
superposition in the central system due to its interaction and subsequent
entanglement with the environmental spins. In particular, one is
interested in the time-dependence of the expectation value of the
$z$-component of the first spin $\langle\sigma_1^z(t)\rangle$,
where $\sigma_1^z$ is the Pauli matrix acting on the state of $\vec c_1$.
In the absence of the coupling to the environment this quantity
exhibits periodic oscillations between $+1$ and $-1$ caused by the
first term in Eq.~(\ref{Model}); coupling to the
environment is expected to damp these oscillations.

In the work reported in Ref.~\cite{DobRaeKanHar} a numerical
investigation of this problem was performed. Among many surprising
features in the behavior of the above system, it was observed that
after an initial fast decay of the oscillations of
$\langle\sigma_1^z(t)\rangle$ the amplitude showed a saturation at
the value of $1/3$. 
Subsequently, the oscillations demonstrate a
much slower decay, which is consistent with the $1/t$ conjecture,
and which leads to a complete suppression of oscillations.
The main motivation of this
paper was to understand the cause of the saturation and the subsequent
slow decay.

While the model Eq.~(\ref{Model}) is hard to treat analytically, we
simplified it by setting all $J_k$'s equal. This allowed us to solve
the model exactly. The solution turned out to reproduce quantitatively
several key features of the numerical results reported
in~\cite{DobRaeKanHar}. In fact, it reproduced the fast initial decay
of the amplitude of oscillations and its subsequent saturation at 1/3.
It also offers a way to qualitatively understand the cause of the
long-time
tail. Most importantly, it answers the question: why is the
decay of oscillations in our model much slower compared to a more
conventional exponential decay of oscillations in, say, the spin-boson
models.~\cite{TLS} The cause is the integer value of total spin of the
central system.

This work shows that integer spins, in contrast to half-integer spins,
may, under suitable circumstances, exhibit quantum oscillations over
much longer times. From the perspective of the theory of quantum phase
transitions, this work also offers a simple example of emergent
power-law correlations usually associated with criticality.


\section{$J_k=J$ model}

To make analytical progress we consider a simplified model where
we take all coupling constants $J_k=J$ to be equal while preserving random
uncorrelated initial states of the environmental spins. The Hamiltonian
takes the form
\begin{eqnarray}
\label{Simplified_model}
H &=& \vec C^2 + 2J\vec C\vec S\nonumber\\
  &=& (1-J)\vec C^2+J(\vec C+\vec S)^2 - J\vec S^2,
\end{eqnarray}
which describes the coupling of the central spin $\vec C=\vec c_1+\vec
c_2$ to the total spin of the environment $\vec S=\sum \vec s_k$.  We
are interested in the expectation value of the $z$-component of $\vec
c_1$: $\langle \sigma_1^z(t)\rangle$, where $\sigma_1^z$ is the Pauli
matrix acting on the state of $\vec c_1$.
Note, that the assumed initial condition Eq.~(\ref{Initial_condition})
corresponds to the
superposition of states with different $S$. The Hamiltonian
Eq.~(\ref{Simplified_model}) conserves $\vec S^2$, therefore the
matrix element $\langle \sigma_1^z(t)\rangle$ can be decomposed as
\begin{eqnarray}
\label{Averaging}
\langle \sigma_1^z(t)\rangle
   = \sum_S \langle S|\sigma_1^z(t)|S\rangle P(S),
\end{eqnarray}
where $P(S)$ is the weight of the state with the total spin $S$ given
the random uncorrelated initial states of $\vec s_k$. We thus are led
to the problem of first calculating $P(S)$.

Before proceeding with the actual calculation an important comment is in
order. Since Eq.~(\ref{Averaging}) looks like an average over all
possible initial orientations of the environmental spins one might interpret
the above quantity $\langle \sigma_1^z(t)\rangle$ as an ensemble-averaged
expectation value. Quite importantly, in the case where the number of
environmental spins is large the actual weight of the state with total
spin of the environment $S$ tends to the ensemble-averaged quantity $P(S)$.
Therefore, in this limit Eq.~(\ref{Averaging}) describes well the evolution
of the central system in a single realization of the experiment.


In the basis where $\vec s_k$ are good quantum numbers the initial density
matrix is, by assumption, a $2^N\times 2^N$ matrix:
\begin{eqnarray}
\rho_i^S = 2^{-N} I,
\end{eqnarray}
where $I$ is a unit matrix.  Let us make a unitary transformation to
the basis spanned by the the eigenstates of $\vec S^2$. There are $N$
different values that $\vec S^2$ can take. To preserve the
dimensionality of the Hilbert space we conclude that some (in fact
almost all) of these latter states are degenerate. A unitary
transformation will leave the initial density matrix unchanged. This
means that
\begin{eqnarray}
P(S) = 2^{-N}G(S)(2S+1),
\end{eqnarray}
where $G(S)$ is the degeneracy of the state with total spin $S$ (with
$S_z$ fixed). To calculate $G(S)$ we change variables and introduce
$g(k)=G(N/2-k)$. The state with the maximum total spin $S=N/2$ is
unique and is the state where all $\vec s_k$'s point up (we choose
$S_z=S$), therefore $g(0)=1$. Next, a state with $S=S_z=N/2-1$ should
be a superposition of the states with $N-1$ spins up and one spin down.
There are $C_1^N=N$ such states ($C_M^N = N!/M!(N-M)!$ is the binomial
coefficient). However, among such states there are $g(0)=1$ states with
$S=N/2$ and $S_z=N/2-1$ which have to be excluded. Generalizing to
arbitrary $k$ we get:
\begin{eqnarray}
g(k) &=& C_k^N - \sum\limits_{i=0}^{k-1}g(i)\nonumber\\
     &=& C_k^N - C_{k-1}^N \nonumber\\
     &=& C_k^N\left(\frac{N-2k+1}{N-k+1}\right).
\end{eqnarray}
We thus have the result for the weight of the state with spin $S$:
\begin{eqnarray}
\label{P_of_S}
P(S) &=& 2^{-N}C_{N/2-S}^N\,\frac{(2S+1)^2}{N/2+S+1}\nonumber\\
     &\approx& \frac{8S^2/N}{\sqrt{2\pi D}}\,e^{-S^2/2D}, \qquad D=N/4,
\end{eqnarray}
where we have used a well-known approximation for the binomial distribution
described by the first two factors above. One can easily check that
$\int_0^\infty P(S)\,dS = 1$, i.e. the approximations we made preserve
the normalization of the probability.


We have thus reduced the problem to finding the time evolution of the
initial state:
\begin{eqnarray}
|f\rangle &=& e^{-iHt}|i\rangle,\\
\label{Initial_state}
|i\rangle &=& \frac{1}{\sqrt{2}} \left( |1,0\rangle_C + |0,0\rangle_C \right)
     |S,S^z\rangle_S,
\end{eqnarray}
estimating the spin polarization $\langle f|\sigma_1^z|f\rangle$, and
averaging the result with respect to $S^z$ (trivial) and $S$ (according
to Eq.~(\ref{P_of_S})). There are two circumstances that greatly simplify
the calculation. First, the Hamiltonian acting on the singlet state
$|0,0\rangle_C$ gives zero, therefore the evolution of the second term
in Eq.~(\ref{Initial_state}) is trivial. Second, the symmetry of the
Hamiltonian with respect to $\vec c_1$ and $\vec c_2$ implies that given
the above initial condition we have
$\langle f|\sigma_1^z|f\rangle = - \langle f|\sigma_2^z|f\rangle$. Thus,
we can calculate the expectation value of
$\sigma_{-}^z=(\sigma_1^z-\sigma_2^z)/2$ instead. For this operator we have:
$\sigma_{-}^z |1,0\rangle_C = |0,0\rangle_C$,
$\sigma_{-}^z |0,0\rangle_C = |1,0\rangle_C$,
$\sigma_{-}^z |1,\pm 1\rangle_C = 0$.
Taking all this into account, we see that
\begin{eqnarray}
\langle f|\sigma_1^z|f\rangle &=& {\rm Re} \, \langle t| e^{-iHt} |t\rangle,\\
   |t\rangle &=& |1,0\rangle_C |S,S^z\rangle_S.
\end{eqnarray}
From Eq.~(\ref{Simplified_model}) it is clear that the above matrix element
can be easily calculated after going to the basis with well defined total
spin $\vec L = \vec C+\vec S$. The necessary Clebsch-Gordan decomposition
(in the limit $S \gg 1$ of interest to us) is:
\begin{eqnarray}
|t\rangle
  &\approx& \sqrt{\frac{1-\left(S^z/S\right)^2}{2}} \,
     (|S+1,S^z\rangle_L - |S-1,S^z\rangle_L) \nonumber\\
  &+& \frac{S^z}{S} \, |S,S^z\rangle_L,
\end{eqnarray}
where we have introduce the notation $|L,L^z\rangle_L$. In this
basis we easily calculate using Eq.~(\ref{Simplified_model}):
\begin{eqnarray}
\langle f|\sigma_1^z|f\rangle
  &=& \cos2(1-J)t \, \times \nonumber\\
  &\times& \left\{ [1-\left(\frac{S^z}{S}\right)^2]\cos 2JSt
    + \left(\frac{S^z}{S}\right)^2\right\}.
\end{eqnarray}
Finally, we have to average this result over $S^z$ and $S$. The first
average is done trivially using the fact that (in the same limit $S\gg 1$)
$\langle (S^z)^2\rangle=S^2/3$. The second average is calculated using
Eq.~(\ref{P_of_S}) which leads to a Gaussian integral. The result is:
\begin{eqnarray}
\label{Result}
\langle \sigma_1^z \rangle
  &=&      A(t) \cos2(1-J)t, \\
A(t) &=& \frac 13 + \frac 23 (1-NJ^2t^2)\,e^{-NJ^2t^2/2}.
\end{eqnarray}
It should be stressed that this result is {\em exact} in the limit $N\gg 1$
($S\gg 1$).
We see that an initial exponential decay of the amplitude of the
oscillations is followed by a transient and an eventual leveling at
$A(t)=1/3$.

\begin{figure}
\label{Numerics}
\includegraphics[width=\columnwidth]{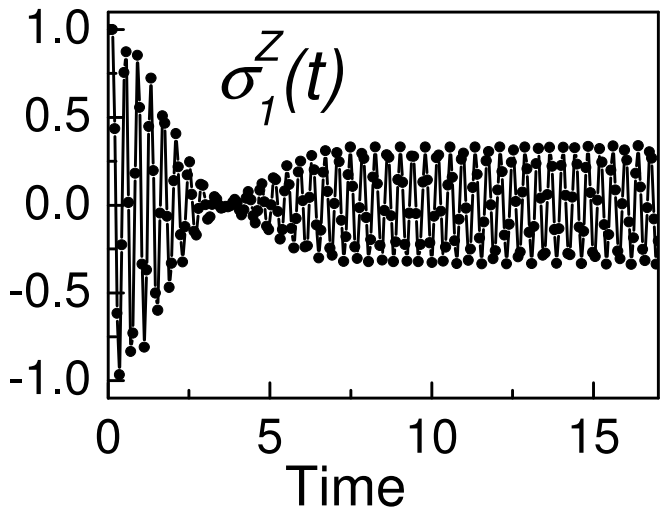}
\caption{Numerical simulation of 13 spins with $J_0 = 8$, $J_k = J =
  0.128$.  The figure shows the expectation value of $\sigma_1^z$
  as a function of time.}
\end{figure}

To check the above results we have performed a direct numerical
solution of the Schr\"odinger equation corresponding of the system
with a Hamiltonian $H = J_0\vec C^2 + 2\vec C\sum_{k=1}^N J_k\vec
s_k$, $J_k=J$, which can be reduced to Eq.~(\ref{Model}) by rescaling
$J_0\to 1$, $J\to J/J_0$, $t\to tJ_0$. Exact diagonalization was used
to find the time evolution. An example of the results is
shown in Fig.~1. It shows the expectation value of $\sigma_1^z$ as a
function of time. The parameters are: the number of spins $N=13$,
$J_0=8$, $J_k=J=0.128$. This can be compared with the analytic result
for the same quantity which is given (after rescaling) by
Eq.~(\ref{Result}) and is shown in Fig.~2. The numerical and
analytical results show excellent agreement.

\begin{figure}
\label{Check}
\includegraphics[width=\columnwidth]{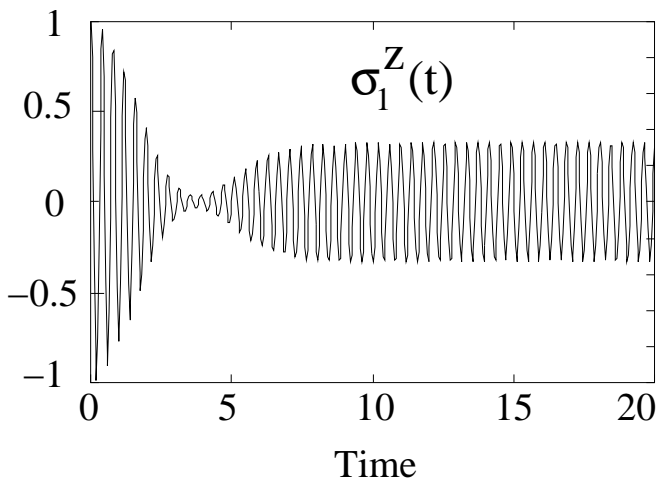}
\caption{Analytical result for $\sigma_1^z(t)$ with the same
parameters as those used in numerical simulations.}
\end{figure}

Absence of the decay of the amplitude of oscillations at long times is
quite an unexpected result. Therefore it is worth explaining it in
some more detail.



\section{Discussion of the results: simple picture}

One trivial situation where the oscillation of the central spin does not
decay is that of no interaction between the central spin and the set of
environmental spins. In the presence of such an interaction, however, one
may still ask what are the conditions under which this interaction is
ineffective in damping the oscillations. A natural suggestion is to try to
find a state $|\Psi\rangle$ of the combined system in which
$\langle \Psi|H_{\rm int}|\Psi\rangle = 0$. Since in our case
$H_{\rm int}=2J\vec C\vec S$, classically such a state $|\Psi\rangle$ would
correspond to vectors $\vec C$ and $\vec S$ being orthogonal. The condition
$\vec C\vec S=0$ defines a plane in 3D space, therefore one could argue
that for the case of random initial orientation of $\vec S$ the probability
of being in the state with $H_{\rm int}=0$ is zero. Remarkably, the quantum
nature of spins proves the result to be quite different.

The correct way of treating $H_{\rm int}$ is, of course, to rewrite it in
the following form:
\begin{equation}
\label{H_int}
H_{\rm int} = J(\vec C + \vec S)^2 - J\vec S^2 - J\vec C^2.
\end{equation}
Adding spin $C=1$ with spin $S$ results in possible values of total spin
$C+S$ being: $S-1$, $S$ and $S+1$. It is the second case in which the first
two terms in the Eq.~(\ref{H_int}) cancel each other. The remaining last
term does not depend on $S$ and, therefore, does not suppress the oscillation
amplitude when the averaging over $S$ is performed and only shifts the
oscillation frequency of the central spin (this effect is reflected in
Eq.~(\ref{Result})). The condition ``$C+S=S$'' is, thus, the closest analog
of the classical condition $\vec C\vec S=0$. But, unlike in the classical
case, a simple Clebsch-Gordan algebra (see previous section)
shows that the probability of being in the subspace ``$C+S=S$'' is
actually finite and is equal to 1/3.

One can easily see now that this effect can only occur if the central system
has integer spin. Indeed, the condition ``$C+S=S$'' can never be satisfied
if $C$ is half-integer. These considerations allow us to formulate the
main result of the paper: Based on a particular model of a central spin
interacting with randomly-oriented environmental spins we have been able to
show that the decay of the oscillations of the central spin is
essentially different for {\it integer} central spins: the decay is no longer
exponential, instead the amplitude of the oscillations saturates at a
constant value.


 
Moreover, the results presented in this work make clear the physical
origin of the unusual two-step decoherence found in
Ref.~\cite{DobRaeKanHar}, where the generic model Eq.~(\ref{Model}) has been
considered with all $J_k$ being different. The first step of
decoherence, associated with the initial decay of oscillations to the
value of 1/3, has been described in Ref.~\cite{DobRaeKanHar} using a
mean-field-like treatment of the spin bath, by replacing the
interaction part of the Hamiltonian with a random classical static
field having Gaussian distribution. However, such a treatment fails to
describe the second step of decoherence, i.e.\ the long-time slow
decay of oscillations.  As the results above demonstrate, the
representation of a bath as a static random field corresponds to the
case of all $J_k$ being equal to $J$. This stems from the fact that
the total spin of the bath $S^2$ commutes with the
Hamiltonian~(\ref{Simplified_model}), so that the bath dynamics in the
case $J_k=J$ is trivial, and can be removed completely by a
transformation into the rotating coordinate system. Then, in the
rotating coordinate system the effect of the bath on the central spins
is equivalent to the action of a random static field. Therefore, the
initial decoherence is similar to the ``adiabatic decoherence'' by a
static spin bath, considered e.g.\ in Ref.~\cite{zurek}.

Correspondingly, the second step of the decoherence process, i.e.\ the
long-time slow decay of quantum oscillations, can be caused only by an
internal evolution of the bath.  For all $J_k$ being different, $S^2$
does not commute with the interaction part of the
Hamiltonian~(\ref{Simplified_model}), and, as a result, the
system-bath coupling induces a non-trivial dynamics inside the bath.
It is not surprising that the spin bath possessing a complex dynamics
can not be represented as a random static magnetic field acting on the
system. Understanding this ``minimally non-adiabatic'' decoherence regime
represents a challenge for future investigations.~\cite{Miyashita_Nagaosa}

Summarizing, in this work we have demonstrated that decoherence of
many-spin systems can drastically differ from decoherence of
single-spin systems.  This difference originates at the most basic
level, and is determined by parity of the central system, i.e.\
whether the system comprises even or odd number of spin-1/2
entities. Therefore, it is very likely that similar distinction
between the central spin systems of even and odd parity is important
in many other situations. Moreover, our consideration clarifies the
origin of the unusual two-step decoherence found numerically in
Ref.~\cite{DobRaeKanHar}.
The exactly solvable model allows clear
demonstration that the initial step of decoherence (associated with
the saturation of oscillations at the value of 1/3) is caused by
``adiabatic decoherence'' by a static spin bath, while the subsequent
long-time slow decay is induced by a non-trivial internal dynamics of
the spin bath.
The model is applicable to the qualitative analysis of
a range of experimental systems such as magnetic molecules and shallow
impurity spins in semiconductors, which experience decoherence from
the nuclear spin bath. In these cases, the dominant interaction with
the nuclear spins is well approximated by the isotropic Heisenberg
interaction (anisotropic interactions are often small).


\begin{acknowledgments}
This work was supported in part by the National Security Agency 
(NSA) and Advanced Research and Development Activity (ARDA) 
under Army Research Office (ARO) contract number 421-25-01.
This work was partially
carried out at the Ames Laboratory, which is operated
for the U. S. Department of Energy by Iowa State
University under Contract No. W-7405-82 and was supported
by the Director of the Office of Science, Office
of Basic Energy Research of the U. S. Department of
Energy. Support from the Dutch
``Stichting Nationale Computer Faciliteiten (NCF)'' is
gratefully acknowledged.

\end{acknowledgments}


\end{document}